\documentclass[pdftex,12pt, pdfminorversion=4]{iopart}
\usepackage{iopams}

\usepackage{amssymb}
\usepackage{bm}
\usepackage[pdftex]{graphicx}
\usepackage[pdftex,breaklinks=true,colorlinks=true]{hyperref}
\usepackage{bm}
\usepackage{braket}

\renewcommand{\mr}[1]{\mathrm{#1}}
\newcommand{\lp}{\left(}
\newcommand{\rp}{\right)}
\newcommand{\delphi}{\partial \phi}

\newcommand{\be}{\begin{equation}}
\newcommand{\ee}{\end{equation}}
\newcommand{\f}{\frac}
\newcommand{\s}{\sqrt}
\newcommand{\p}{\partial}

\newcommand{\bea}{\begin{eqnarray}}
\newcommand{\eea}{\end{eqnarray}}
\newcommand{\ba}{\begin{align}}
\newcommand{\ea}{\end{align}}

\newcommand{\la}{\langle}
\newcommand{\ra}{\rangle}

\begin{document}

\title{Numerical calculations on the relative entanglement entropy in critical spin chains }

\author{Yuya O. Nakagawa$^1$ \& Tomonori Ugajin$^2$}

\address{$^1$ Institute for Solid State Physics, the University of Tokyo, Kashiwa, Chiba 277-8581, Japan}
\ead{y-nakagawa@issp.u-tokyo.ac.jp}
\address{$^2$ Kavli Institute for Theoretical Physics,  University of California, Santa Barbara, CA 93106, USA}
\ead{ugajin@kitp.ucsb.edu}

\vspace{10pt}
\begin{indented}
\item[] \today
\end{indented}

\begin{abstract}
We study the relative entanglement entropy (EE) among various primary excited states
in two critical spin chains: the $S=1/2$ XXZ chain and the transverse field Ising chain at criticality.
For the $S=1/2$ XXZ chain, which corresponds to $c=1$ free boson conformal field theory (CFT),
we numerically calculate the relative EE by exact diagonalization
and find a perfect agreement with the predictions by the CFT.
For the transverse field Ising chain at criticality, which corresponds to the $c=1/2$ Ising CFT,
we analytically relate its relative EE to that of the $S=1/2$ XXZ chain and confirm the relation numerically.
We also calculate the ``sandwiched" R\'{e}nyi relative EE and again the numerical results agree well with the 
analytical predictions. Our results are the first direct confirmation of the CFT predictions on the relative EE of the primary excited states
in critical spin chains. 
\end{abstract}

\section{ Introduction} 
Entanglement entropy quantifies the amount of entanglement of pure states,  and  is defined by von Neumann entropy of the reduced density matrix. It has been used to study various quantum many-body systems, such as (1+1)-dimensional
conformal field theories (CFTs)~\cite{Holzhey:1994we, Calabrese:2004eu, Calabrese:2009qy} or
(2+1)-dimensional topological phases~\cite{Kitaev:2005dm, 2006PhRvL..96k0405L},
and also naturally appears in AdS/CFT correspondence~\cite{Maldacena:1997re} which connects a class of large $N$ gauge theories in $d$ dimensions to classical theories of gravity in $(d+1)$-dimensional anti-de~Sitter space ~\cite{Ryu:2006bv,Ryu:2006ef}.
It was also argued that entanglement is strongly related to the spacetime structure itself,  dubbed ER=EPR~\cite{Maldacena:2013xja}.

Relative entropy $S( \rho|| \sigma)$ is a useful distance measure between two density matrices 
$\rho$ and $\sigma$~\cite{araki1976relative}. It is defined by 
\be
S( \rho|| \sigma)= {\rm tr} \lp \rho \log \rho \rp - {\rm tr} \lp \rho \log \sigma \rp \label{Def_RelEE},
\ee
and has several nice properties~\cite{RevModPhys.50.221}.
First of all,  this quantity is positive definite and becomes zero only when $\rho= \sigma$. 
Second, when we consider reduced density matrices of some region $A$ in a quantum field theory,
although their von Neumann entropies suffer from the ultraviolet divergence coming from short-distance entanglement,
the relative entropy between them is finite and therefore remains to be well-defined.
Furthermore the relative entropy is free from the ambiguity of the choice of the operator algebra of the region in question \cite{Casini:2013rba}.
Third,  it satisfies the monotonicity property: when the region $A$ is included in the region $B$, $A \subset B$,
the relative entropy of the region $B$ is larger than that of the region $A$,
$ S (\rho_{A}|| \sigma_{A}) \leq S (\rho_{B}|| \sigma_{B})$, where $\rho_{A(B)}, \sigma_{A(B)}$ are the
reduced density matrices of the states $\rho, \sigma$ for the region $A(B)$.
These properties play a crucial role in the recent development of information theoretic approach to quantum field theories, see for example~\cite{Faulkner:2016mzt, Casini:2016fgb,Casini:2016udt,Bousso:2014sda,Casini:2017vbe}.  \footnote{ A nice review of relative entropy is~\cite{Vedral:2002zz}.}

It is nevertheless often difficult to  compute  relative entropy between two reduced density matrices in quantum field theories.
In~\cite{Lashkari:2015dia}
a replica trick to calculate the relative entropy was introduced.
Based on this trick, a general formula for the relative entropy between two arbitrary reduced density matrices
of small subsystems in (1+1)-dimensional CFT was derived~\cite{Sarosi:2016oks}.
The result was later generalized to higher-dimensional CFT~\cite{Sarosi:2016atx, Sarosi:2017rsq}
and the case of two disjoint subsystems~\cite{Ugajin:2016opf}.
When one of the reduced density matrices is coming from the vacuum, one can compare these results
with the holographic ones derived from the holographic entanglement entropy formula ~\cite{Ryu:2006bv,Ryu:2006ef},
and the calculations in the gravity side completely agree with the  CFT results~\cite{Sarosi:2016atx}.~\footnote{There are other holographic studies of the relative entropy, for example ~\cite{Lin:2014hva, Jafferis:2015del, Lashkari:2016idm}.} Also, in \cite{Caputa:2016yzn} the dynamics of the relative entropy between
the locally excited states and the ground state in (1+1)-dimensional CFT was calculated.

In \cite{Ruggiero:2016khg} Ruggiero and Calabrese studied the R{\'e}nyi relative entropy,
\be
S_{n} (\rho|| \sigma) = \frac{1}{1-n} \log \lp \frac{ \tr \rho \sigma^{n-1}}{\tr \rho^n}  \rp,
\:\: \lim_{n \to 1} S_{n} (\rho|| \sigma) = S (\rho \| \sigma),  \label{Def_RenyiRelEE}
\ee
between several excited states in the XX spin chain model and compared the results with the CFT predictions.
They used the formula which relates the correlation matrix of the system to
the R\'enyi relative entropy $S_n$ in numerical calculations, and
the numerical results match the CFT predictions well.
This method, however, is only applicable to free (quadratic) systems and
it is difficult to take the limit $n \rightarrow 1$, or calculate the relative entropy itself.
In addition, the definition of the R{\'e}nyi relative entropy~\eref{Def_RenyiRelEE}
is different from the one conventionally used in the field of quantum information theory,
and its meaning from the viewpoint of quantum information theory has not been obvious yet.

In this paper we perform a numerical study of the relative entropy itself
rather than the R\'enyi counterpart of it in critical spin chain models.
In particular, we consider the $S=1/2$ XXZ chain
and the critical transverse field Ising chain under the periodic boundary condition.  
We utilize exact diagonalization to explicitly construct reduced density matrices of excited states of the models 
and calculate relative entropies among them. 
The numerical results show a perfect agreement
with the analytical results of the corresponding CFTs
($c=1$ free boson CFT for the XXZ chain and the $c= 1/2$ Ising CFT for the critical transverse field Ising chain).
Moreover, by taking advantage of exact diagonalization,
we also calculate the ``sandwiched" R\'enyi relative entropy~\cite{Lashkari:2014yva,Sandwich2013, Wilde2014,2013JMP....54l2201F}
of the order of $\alpha =1/2$, or quantum fidelity, between excited eigenstates of the XXZ chain.
The numerical results again match the CFT predictions in~\cite{Lashkari:2014yva}.

This paper is organized as follows. In section~\ref{section:def} we explain our setup
and how to compute the relative entropy by using the replica trick~\cite{Lashkari:2015dia}
in two-dimensional CFT.
In section~\ref{section:c=1} we review the previous results of the relative entropy in $c=1$ CFT~\cite{Lashkari:2015dia,Ruggiero:2016khg}.
The sections~\ref{section:c=1/2} and \ref{section:numerics} consist of our new results.    
In section~\ref{section:c=1/2} we derive analytic expressions of the relative entropy between several
primary excited states in the $c= 1/2$ Ising CFT.
In section~\ref{section:numerics} we present extensive numerical results
on the relative entropy in critical spin chains.
In section \ref{section:conc}, we conclude our study and discuss future work.

\section{ Basic definitions} \label{section:def}
\subsection{Setup}
We start from a two-dimensional CFT on a cylinder $\mathbb{R}_{t} \times S^{1}$  with the coordinates $(t, \phi), \phi \sim \phi +2 \pi$. We choose our subsystem $A$ to be the segment
$\{ \phi |  -\pi x<\phi < \pi x\}$ at $t=0 \:\: (0<x<1)$,
and consider the reduced density matrix $\rho_{V}$ of an excited state $| V \ra$, 
\be
\rho_{V} = {\rm tr}_{A_{c}} | V \ra \la V|, 
\ee 
where $A_{c}$ denotes the complement of the region $A$.

The quantity ${\rm tr} \rho_{V}^{n} $
can be calculated by a path integral on the $n$-sheet cylinder with the cut along the subsystem $A$
and the boundary conditions at $ t= \pm \infty$ specifying the excited state $|V \ra$.  
Furthermore, by applying a uniformarization map and using the state-operator correspondence,
we see that the path integral is proportional to the $2n$ point function on the cylinder~\cite{Alcaraz:2011tn, Berganza:2011mh},
\be
\fl
F^{n}_{V} (x)  \equiv  \f{{\rm tr} \rho_{V}^{n}}{{\rm tr} \rho_{(0)}^{n}} = \left(\f{1}{n} \right)^{2n\Delta_{V}}\frac{\tilde{F}^{n}_{V}(x)}{ \la V(t_{0,1})  V^{\dagger}(t'_{0,1}) \ra^{n}}, \quad \tilde{F}^{n}_{V}(x) \equiv \la \prod^{n-1}_{k=0} V(t_{k,n}) V^{\dagger} (t'_{k,n})\ra \label{eq:fv}
\ee
where $\frac{1}{1-n} \log \lp {\rm tr} \rho_{(0)}^{n} \rp$
is the $n$-th R{\' e}nyi entanglement entropy of the vacuum,
the $V(z)$ is the  local operator corresponding  to the excited state~$\ket{V}$ with the conformal dimensions $(h_{V},\bar{h}_{V}), \; \Delta_{V}= h_{V}+\bar{h}_{V}$,
and  $(t_{k,n}, t'_{k,n}) $ are given by 
\be
t_{k,n} =\frac{\pi}{n} (x+2k),  \quad t'_{k,n} =\frac{\pi}{n} (-x+2k).
\ee
Also, $V^{\dagger}(z)$ denotes the
Belavin-Polyakov-Zamolodchikov conjugate of $V$.

\subsection{Replica tricks for relative entropy} 
To compute the relative entropy $S( \rho || \sigma)$ between two density matrices $\rho$ and $\sigma$,
it is useful to introduce the following replica trick,
\be
\log G_n (\rho|| \sigma) =  \log \tr \rho^{n}- \log \tr \rho \sigma^{n-1}, \label{replica}
\ee
which recovers the relative entropy in the limit of $n \rightarrow 1 $, 
\be
S(\rho|| \sigma) = \lim_{n \rightarrow 1} \f{1}{n-1} \log G_{n}(\rho|| \sigma). \label{replica_limit}
\ee

We are interested in the case where $\rho$ and $\sigma$ are reduced density matrices of excited states
$|V \ra, | W \ra$, $\rho= \rho_{V},\;  \sigma = \rho_{W} \equiv {\rm tr}_{A_{c}} |W \ra \la W|$. 
In this setup the second term of~\eref{replica} also can be written
in terms of the correlation function of $V$ and $W$, 
\begin{eqnarray}
\fl
F^{1,n-1}_{V,W}(x)& \equiv& \f{{\rm tr} \rho_{V}\rho_{W}^{n-1}}{{\rm tr} \rho_{(0)}^{n}} \nonumber \\[+10pt]
&=& \left(\f{1}{n} \right)^{2\Delta_{V}+2(n-1)\Delta_{W}}\frac{\tilde{F}^{1,n-1}_{V,W}(x)}{\la V(t_{0,1})  V^{\dagger}(t'_{0,1}) \ra \la W(t_{0,1})  W^{\dagger}(t'_{0,1}) \ra^{n-1}} \label{eq:fvw}
\end{eqnarray}  
with 
\be
\tilde{F}^{1,n-1}_{V,W}(x)\equiv\la V(t_{0,n}) V^{\dagger} (t'_{0,n})\prod^{n-1}_{k=1} W(t_{k,n}) W^{\dagger} (t'_{k,n})\ra
\ee
where $\Delta_{W}$ is the scaling dimension of the operator $W$.
In terms of these quantities
the relative entropy between two reduced density matrices $\rho_{V},\rho_{W}$ is given by 
\be
\log G_{n}(\rho_{V}|| \rho_{W}) =\log F^{n}_{V} (x) - \log F^{1,n-1}_{V,W}(x),
 \label{eq:renrelative}
\ee
and \eref{replica_limit}.

Another replica trick known in the literature is
the sandwiched R\'enyi relative entropy~\cite{2013JMP....54l2201F, Lashkari:2014yva} defined by 
\be
\hat{S}_{\alpha} (\rho|| \sigma) =\f{1}{\alpha-1} \log {\rm tr} (\rho_{\alpha})^{\alpha}, \qquad
\rho_{\alpha}= \sigma^{\f{1-\alpha}{2\alpha}} \rho \sigma^{\f{1-\alpha}{2\alpha}}. \label{eq:sand}
\ee
In the limit of $\alpha \rightarrow 1$, (\ref{eq:sand}) yields the relative entropy~\eref{Def_RelEE}.
The sandwiched R\'enyi relative entropy has several nice properties from the viewpoint of quantum information theory
such as monotonicity under completely positive, trace-preserving maps ~\cite{2013JMP....54l2201F}.  
Also when $\alpha =1/2$,
this is related to another measure of distance of density matrices called quantum fidelity $F(\rho|| \sigma)$ ~\cite{nielsen2010quantum}, 
\be 
 F(\rho|| \sigma) \equiv {\rm tr} \s{\s{\sigma}\rho\s{\sigma}}
 = \exp \left[ -\frac{1}{2} \hat{S}_{\f{1}{2}} (\rho|| \sigma) \right].
\ee

\section{ Relative entropy in $c=1$ CFT} \label{section:c=1}
In this section we discuss relative entropy in $c=1$ CFT. 
The results in this section have already derived in~\cite{Ruggiero:2016khg} , thus the purpose of this section is briefly summarizing them for the  comparisons with numerical results in the later sections. 
  
$c=1$ CFT  is the theory of a  free real boson. The action is given by 
\be
S=\f{g}{8\pi} \int dz^2 \p \varphi \bar{\p} \varphi.  \label{freeboson_action}
\ee
For the later purpose it is convenient to introduce chiral and anti-chiral fields $ \phi(z), \bar{\phi}(\bar{z}), \; \varphi = \phi(z)- \bar{\phi}(\bar{z})$.
There are two primary operators of interest in the theory. 
The first one is the (non-chiral) vertex operator defined by
\be
V_{\alpha,\alpha}(z, \bar{z})= :e^{i\alpha (\phi(z)+ \bar{\phi} (\bar{z})) }:, \label{eq:vertexop}
\ee
and its conformal dimensions are 
\be
\left( h_{\alpha}, \bar{h}_{\alpha} \right) = \left( \f{\alpha^2}{2g}, \; \f{\alpha^2}{2g} \right).
\ee
The second one is the current operator $ i \p \phi$ whose conformal dimensions are 
\be
\left(h_{i\p \phi},\; \bar{h}_{i\p \phi} \right) = \left(1,0 \right).
\ee

\subsection{$S(\rho_{V_{\alpha, \alpha}}|| \rho_{GS}) $ \label{vertex_GS}} 
Let us begin with the relative entropy $S(\rho_{V_{\alpha, \alpha}}|| \rho_{GS})$ between the reduced density matrix $\rho_{V_{\alpha, \alpha}}$ of the vertex operator $V_{\alpha, \alpha}$ and that
of the ground state, $\rho_{GS}$. Now \eref{eq:renrelative}  is given by 
\be
\log G_{n}(\rho_{V_{\alpha, \alpha}}|| \rho_{GS}) =  \log F^{n}_{V_{\alpha, \alpha} } (x) - \log  F^{1,n-1}_{V_{\alpha, \alpha} , 1}.
\ee
It was shown   in \cite{Alcaraz:2011tn,Berganza:2011mh} that the first term is independent of $x$,  $F^{n}_{V_{\alpha, \alpha} } (x) =1$.
The second term is universal (fixed solely by conformal symmetry),
\be
F^{1,n-1}_{V_{\alpha, \alpha} , 1} = \left(\f{1}{n}\right)^{2\Delta_{\alpha}} \frac{\la V_{\alpha, \alpha}(t_{0,n}) V^{\dagger}_{\alpha, \alpha}(t'_{0,n})\ra}{\la V_{\alpha, \alpha}(t_{0,1}) V^{\dagger}_{\alpha, \alpha}(t'_{0,1})\ra} =\left( \f{\sin \pi x}{n\sin \f{\pi x}{n}} \right)^{2\Delta_{\alpha}}, \quad \Delta_{\alpha}= h_{\alpha}+\bar{h}_{\alpha}. \label{eq:fnn}
\ee
From these expressions we reach
\be
S(\rho_{V_{\alpha, \alpha}}|| \rho_{GS}) = 2\Delta_{\alpha}  (1-\pi x \cot \pi x). \label{formula_vertex_gs}
\ee
Also, the argument here makes it clear that for any state $V$ with
vanishing R{\'e}nyi entropy $ F^{n}_{V}(x) =1$, the relative entropy is given by
\be
S(\rho_{V}|| \rho_{GS}) = 2 \Delta_{V}  (1-\pi x \cot \pi x), \label{formula_general_V_gs}
\ee
where $\Delta_{V}$ is the scaling dimension of the state.

\subsection{$S(\rho_{i\p \phi} || \rho_{GS})$}
Next we consider the relative entropy $S(\rho_{i\p \phi} || \rho_{GS})$  between the reduced density matrix $\rho_{i\p \phi}$ of the current operator  and $\rho_{GS}$. 
In this case, the first term of  (\ref{eq:renrelative})  is non-trivial,
\be
\fl
F^{n}_{i\p \phi}(x) = \left(\f{1}{n}\right)^2 \f{\la \prod^{n-1}_{k=0} i\p \phi (t_{k,n}) i\p \phi (t'_{k,n}) \ra}{\la  i\p \phi (t_{0,1}) i\p \phi (t'_{0,1}) \ra^{n}} =\frac{1}{n^2\left[4\la  i\p \phi (t_{0,1}) i\p \phi (t'_{0,1}) \ra \right]^{n}} {\rm det} \left[ \f{1}{\sin \f{t_{ij}}{2}}\right]_{i,j \in [1,2n]}. \label{eq:renyip}
\ee
$t_{ij}$ in the right-hand side is defined by 
\be
t_{2k+1}=t_{k,n}, \quad t_{2k+2} =t'_{k,n}, \quad t_{ij}=t_{i}-t_{j}.
\ee
The analytic continuation of the determinant was worked out in~\cite{2013PhRvL.110k5701E,2014JSMTE..09..025C},  and the result is 
\be
{\rm det} \left[ \f{1}{\sin \f{t_{ij}}{2}}\right]_{i,j \in [1,2n]} = 4^{n} \frac{\Gamma^2 (\frac{1+n+n\csc \pi x}{2})}{\Gamma^2 (\frac{1-n+n\csc \pi x}{2})}. 
\ee
The second term of the replica (\ref{eq:renrelative})  in this case is again universal as in (\ref{eq:fnn}), and by taking $n \rightarrow 1$ limit, we obtain
\be
S(\rho_{i\p \phi} || \rho_{GS})=2 \left( \log (2 \sin \pi x) +1-\pi x\cot \pi x +\psi_{0} \left(\f{\csc \pi x}{2} \right) + \sin \pi x \right), \label{formula_delphi_gs}
\ee
where $\psi_{0}(x)$ is the digamma function.

\subsection{ $S(\rho_{i \p \phi}|| \rho_{V_{\alpha, \alpha}})$ \label{delphi_vertex_analytical} }
Here we consider the relative entropy $S(\rho_{i \p \phi}|| \rho_{V_{\alpha, \alpha}})$ between the current operator and the vertex operator. 
The first term of \eref{eq:renrelative} has already computed in the previous subsection. 
The second term  is given by 
\be
\tilde{F}^{1,n-1}_{ i\p \phi, V_{\alpha, \alpha}}(x)=\la i \p \phi (t_{0,n}) i\p \phi(t'_{0,n})\prod^{n-1}_{k=1} V_{\alpha, \alpha} (t_{k,n}) V^{\dagger}_{\alpha, \alpha} (t'_{k,n}) \ra.
\ee
It was argued~\cite{Ruggiero:2016khg} that in the limit of $n \rightarrow 1$ 
this correlator effectively gets factorized  
\begin{eqnarray}
\tilde{F}^{1,n-1}_{ i\p \phi, V_{\alpha, \alpha}}(x)&=&\la i \p \phi (t_{0,n}) i\p \phi(t'_{0,n}) \ra \la \prod^{n-1}_{k=1} V_{\alpha, \alpha} (t_{k,n}) V^{\dagger}_{\alpha, \alpha} (t'_{k,n}) \ra +O((n-1)^2) \nonumber \\[+10pt]
&=&\tilde{F}^{1,n-1}_{i \p\phi, 1}(x) \tilde{F}^{1,n-1}_{1, V_{\alpha, \alpha}} (x) +O((n-1)^2),
\end{eqnarray}
and with the definitions of $F^{n}_{V}(x), F^{1,n-1}_{V,W}(x)$ (\ref{eq:fv}, \ref{eq:fvw}), we have 
\begin{eqnarray}
\log G_{n}(\rho_{i \delphi}|| \rho_{V_{\alpha, \alpha}}) &=&\log F^{n}_{i \p\phi} (x) - \log  F^{1,n-1}_{i \p\phi, V_{\alpha, \alpha}}(x) \nonumber \\[+10pt]
&=& \log G_{n} (\rho_{i \p \phi}|| \rho_{GS}) + \log G_{n}(\rho_{GS}|| \rho_{V_{\alpha, \alpha}}), \quad n\rightarrow 1.
\end{eqnarray}
Therefore the relative entropy $S(\rho_{i \p \phi}|| \rho_{V_{\alpha, \alpha}}) $ is given by the sum of the two relative entropies, 
\be
S(\rho_{i \p \phi}|| \rho_{V_{\alpha, \alpha}}) = S (\rho_{i \p \phi}|| \rho_{GS})+S(\rho_{GS}|| \rho_{V_{\alpha, \alpha}}).  \label{relation_delphi_vertex}
\ee

\section{ Relative entropy in $c= 1/2$ CFT \label{section:c=1/2} }
In this section we consider the relative entropy in the $c= 1/2$ Ising CFT and
show analytical results of  it for the first time.
We also
reveal the relationship between the relative entropies in $c=1$ free boson CFT
and the $c=1/2$ Ising CFT.

$c= 1/2$ Ising CFT is the theory of a free Majorana fermion. The action is given by 
\be
S= \f{1}{8\pi} \int dz^2 \left[\psi \p \psi + \bar{\psi} \bar{\p} \bar{\psi} \right].  \label{c=1/2: action}
\ee
In the discussion below, we focus on two primary operators in this theory,
namely the spin operator $\sigma$ and the energy operator $\varepsilon$.
The conformal dimensions of them are 
\be
\left(h_{\sigma}, \bar{h}_{\sigma} \right) =\left( \f{1}{16}, \f{1}{16} \right) , \quad \left(h_{\varepsilon}, \bar{h}_{\varepsilon} \right) =\left( \f{1}{2}, \f{1}{2} \right).
\ee

It is useful to introduce the bosonization technique to calculate correlation functions of these operators.
Let us consider two copies of the $c=1/2 $ Ising CFT,~i.e., the theory of  two decoupled Majorana fermions $\psi_{1}, \psi_{2}$.
One can form a Dirac fermion $D(z)$ from them \cite{DiFrancesco:1997nk}, 
\be
D (z) = \f{1}{\s{2}}\left( \psi_{1} +i \psi_{2} \right), \quad \bar{D}( \bar{z}) = \f{1}{\s{2}}\left(\bar{\psi}_{1}+i  \bar{\psi}_{2} \right).
\ee
The Dirac fermion can be bosonized, 
\be
D(z) =e^{i\phi(z)} , \quad \bar{D}( \bar{z}) =e^{i \bar{\phi}(\bar{z})},
\ee
where $\phi, \bar{\phi}$ consist of holomorphic and anti-holomorphic part of the boson field $\varphi$,
$\varphi(z, \bar{z}) \equiv \phi(z)-\bar{\phi}(\bar{z})$. 
This boson field $\varphi$ is related to products of the primary operators in two copies of the Ising CFT,
such as
\be
\varepsilon_1 = i \psi_1 \bar{\psi}_1, \:  \varepsilon_2 = i \psi_2 \bar{\psi}_2 
\; \rightarrow \; \varepsilon_{1} \varepsilon_{2} = \p \varphi \bar{\p}\varphi, 
\ee
and
\be
\sigma_{1}\sigma_{2} =\sqrt{2}\cos \f{\varphi}{2}.
\ee
We will use these properties in calculations of the relative entropies in the Ising CFT.

\subsection{$S(\rho_{\sigma}||\rho_{GS})$}
Let us consider the relative entropy between the spin operator $\sigma$ and the ground state. 
The replica trick~\eref{eq:renrelative} is now
\be
\log G_{n}(\rho_{\sigma}||\rho_{GS}) = \log F^{n}_{\sigma}(x) -\log F^{1,n-1}_{\sigma, 1} (x). \label{eq:spings}
\ee
The first term can be computed by the bosonization explained in the previous subsection~\cite{Alcaraz:2011tn,Berganza:2011mh}
\begin{eqnarray}
\la \prod^{n-1}_{k=0}\sigma (t_{k,n}) \sigma  (t'_{k,n})\ra^2  &=& \la \prod^{n-1}_{k=0}\sigma_{1} (t_{k,n}) \sigma_{2} (t_{k,n})\sigma_{1}  (t'_{k,n})\sigma_{2}  (t'_{k,n})\ra \nonumber \\
&=&2^{n}\la  \prod^{n-1}_{k=0} \cos \f{\varphi }{2}(t_{k,n}) \cos \f{\varphi}{2} (t'_{k,n}) \ra.
\end{eqnarray}
One can show that this correlation function is identically constant, $ F^{n}_{\sigma}(x) =1$~\cite{Berganza:2011mh}.  
Therefore, the non-trivial part of  the relative entropy comes from
the second term of  \eref{eq:spings},
and by following the discussions in section~\ref{vertex_GS} we obtain
(note that the scaling dimension of $\sigma$ is $\Delta_\sigma = h_\sigma + \bar{h}_\sigma = 1/8$)
\be
S(\rho_{\sigma}|| \rho_{GS}) = \f{1}{4} \left( 1-\pi x \cot \pi x \right). \label{TI_sigma_gs}
\ee

\subsection{ $S(\rho_{ \varepsilon}||\rho_{GS})$}
Next we discuss the relative entropy between the energy operator $\varepsilon$ and the ground state. 
Again, \eref{eq:renrelative} is written as
\be
\log G_{n}(\rho_{\varepsilon}||\rho_{GS}) = \log F^{n}_{\varepsilon}(x) -\log F^{1,n-1}_{\varepsilon, 1} (x).
\ee
By the bosonization technique
one can relate  the first term to $F^{n}_{i\p \phi } (x) $ computed in the previous section (\ref{eq:renyip}),
\begin{eqnarray}
(\tilde{F}^{n}_{\varepsilon}(x))^2 =\la \prod^{n-1}_{k=0}\varepsilon (t_{k,n}) \varepsilon  (t'_{k,n})\ra^2  &=& \la  \prod^{n-1}_{k=0} \p \varphi \bar{\p}\varphi (t_{k,n}) \p \varphi \bar{\p}\varphi (t'_{k,n}) \ra \nonumber \\
&=& \left| \tilde{F}^{n}_{i\p \phi } (x) \right|^2. 
\end{eqnarray}
Similarly we have $ |\tilde{F}^{1,n-1}_{\varepsilon, 1} (x)| =\tilde{F}^{1,n-1}_{i\p \phi, 1} (x)$.

From these observations we conclude that the relative entropy $S(\rho_{ \varepsilon}||\rho_{GS})$
interestingly coincides with the one between the current operator and the ground state in $c=1$ CFT,
\be
S(\rho_{\varepsilon}|| \rho_{GS}) =S(\rho_{i\p \phi }|| \rho_{GS}). \label{TI_epsilon_GS}
\ee

\subsection{ $S(\rho_{ \varepsilon}||\rho_{\sigma})$}
Finally we consider the relative entropy between
the energy operator $ \varepsilon$ and the spin operator $\sigma$. 
In this case we need to compute 
\be
\tilde{F}^{1,n-1}_{\varepsilon, \sigma}(x)= \la \varepsilon(t_{0,n})  \varepsilon(t'_{0,n}) \prod^{n-1}_{k=1} \sigma (t_{k,n}) \sigma(t'_{k,n}) \ra.
\ee
By the bosonization one can write 
\be 
\tilde{F}^{1,n-1}_{\varepsilon, \sigma}(x)^2 =\sum_{\{\beta_{k}, \bar{\beta}_{k}\}=\pm \f{1}{2}} \f{1}{2^{n-1}}\la \p \varphi \bar{\p}\varphi (t_{0,n})  \p \varphi \bar{\p}\varphi (t'_{0,n}) \prod^{n-1}_{k=1} V_{\beta_{k}, -\beta_{k}}(t_{k,n})  V_{\bar{\beta}_{k}, -\bar{\beta}_{k}}(t'_{k,n}) \ra,
\ee
where $V_{\beta, \gamma}=:e^{i(\beta \phi + \gamma\bar{\phi})}:$.
Each term of  the right-hand side is factorized into holomorphic and anti-holomorphic parts,
\begin{eqnarray}
\fl
\la \p \varphi \bar{\p}\varphi (t_{0,n})  \p \varphi \bar{\p}\varphi (t'_{0,n}) \prod^{n-1}_{k=1} V_{\beta_{k}, -\beta_{k}}(t_{k,n})  V_{\bar{\beta}_{k}, -\bar{\beta}_{k}}(t'_{k,n}) \ra &=& \la \p \varphi  (t_{0,n}) \p \varphi  (t'_{0,n})
\prod^{n-1}_{k=1} V_{\beta_{k},0} (t_{k,n})  V_{\bar{\beta}_{k},0} (t'_{k,n}) \ra \nonumber \\
&\times & \la \bar{\p}\varphi (\bar{t}_{0,n})  \bar{\p}\varphi (\bar{t'}_{0,n}) \prod^{n-1}_{k=1} V_{0,-\beta_{k}} (\bar{t}_{k,n}) V_{0,-\bar{\beta}_{k}} (\bar{t'}_{k,n}) \ra \nonumber.
\end{eqnarray} 
As we have seen in section \ref{delphi_vertex_analytical},
those correlators in the right-hand side are further factorized in $n \rightarrow 1$ limit, 
\begin{eqnarray}
&\left< \p \varphi  (t_{0,n}) \p \varphi  (t'_{0,n})
\prod^{n-1}_{k=1} V_{\beta_{k},0} (t_{k,n})  V_{\bar{\beta}_{k},0} (t'_{k,n}) \right> \nonumber  \\
&= \la \p \varphi  (t_{0,n}) \p \varphi  (t'_{0,n}) \ra \la \prod^{n-1}_{k=1} V_{\beta_{k},0} (t_{k,n})  V_{\bar{\beta}_{k},0} (t'_{k,n}) \ra, \quad n \rightarrow 1.
\end{eqnarray}
This means that 
\begin{eqnarray*}
\tilde{F}^{1,n-1}_{\varepsilon, \sigma}(x)^2 &=& \la \p \varphi \bar{\p}\varphi (t_{0,n})  \p \varphi \bar{\p}\varphi (t'_{0,n}) \ra \sum_{\{\beta_{k}, \bar{\beta}_{k}\}=\pm \f{1}{2}}\f{1}{2^{n-1}} \la \prod^{n-1}_{k=1} V_{\beta_{k}, -\beta_{k}}(t_{k,n})  V_{\bar{\beta}_{k}, -\bar{\beta}_{k}}(t'_{k,n}) \ra \\
&=&\left|\tilde{F}^{1,n-1}_{\varepsilon,GS}(x) \right|^2 \left|\tilde{F}^{1,n-1}_{GS, \sigma}(x) \right|^2,
\:\:\: n \rightarrow 1.
\end{eqnarray*} 
Hence we reach the relation 
\be
S(\rho_{\varepsilon}||\rho_{\sigma}) = S(\rho_{\varepsilon}|| \rho_{GS})+ S(\rho_{GS}||\rho_{\sigma}),
\label{TI_epsilon_sigma}
\ee
which is similar to the case of $c=1$ CFT~\eref{relation_delphi_vertex}.

\section{Numerical results \label{section:numerics}}
In this section, we show numerical results of the relative entropies in critical spin chains.
We first present results for the $S=1/2$ XXZ chain which corresponds to $c=1$ free boson CFT
and then show results for the critical transverse field Ising chain which corresponds to the $c=1/2$ Ising CFT.
For both systems, the analytical predictions by the CFTs described in the previous sections
agree perfectly with the numerical data.

\subsection{$S=1/2$ XXZ chain: $c=1$ free boson CFT}
The first example we examine here is the $S=1/2$ XXZ chain. The Hamiltonian is given by
\begin{equation}
 H = \sum_{i=1}^L \left( S^x_i S^x_{i+1}+S^y_i S^y_{i+1}+  \Delta S^z_i S^z_{i+1} \right),
\label{XXZ_model}
\end{equation}
where $S^{x,y,z}_i $ denotes a $S=1/2$ spin on the site $i$ and
periodic boundary condition is imposed.
The low-energy physics of this model for $|\Delta| \leq 1$ is known to be well described by
a free boson field theory~\eref{freeboson_action} with $g = 4/K$~\cite{GiamarchiBook},
where $K$ is called the Luttinger parameter which
dictates the low-energy properties of the system.
In the case of the XXZ chain, $K$ is exactly known as $K = \pi/(2 \cos^{-1}(-\Delta))$.

In the following, we compare numerical results of the relative entropies for the $S=1/2$ XXZ chain
with the predictions by $c=1$ CFT in section~\ref{section:c=1}.
To this end, we perform exact diagonalization of the model~(\ref{XXZ_model})
and compute energy eigenstates
corresponding to the primary states of $c=1$ CFT.
From those eigenstates, we numerically construct the reduced density matrices
for the subregion $A$ composed of $l$ consecutive sites
and calculate the relative entropies among them.

Before showing the numerical results,
we introduce a dual field of $\varphi$~\cite{GiamarchiBook}, $\theta \equiv (\phi + \overline{\phi})/K $ and
a vertex state of $\theta$, $V_\theta = :e^{i\theta}:$.
This is a specific example  of generic vertex operators $V_{\alpha, \alpha}$ defined in~(\ref{eq:vertexop}),
i.e. $\alpha =\f{1}{K}$. 
In this section we focus on this vertex operator $V_{\theta}$ 
 because it turns out to be numerically easier and stabler
to compute the eigenstate corresponding to $V_\theta$ than other vertex states.    
The conformal dimensions of $V_\theta$ are $(h_\theta, \bar{h}_\theta) = ( 1/(8K), 1/(8K) )$
and the scaling dimension is $\Delta_\theta = 1/(4K)$.
The properties of correlation functions of $V_\theta$ are basically the same as
those of $V_{\alpha, \alpha}$, and the relative entropies involving $V_\theta$ 
are the same as those of $V_{\alpha, \alpha}$, (\ref{formula_vertex_gs},~\ref{relation_delphi_vertex}), 
when the scaling dimension $\Delta_\alpha$ is replaced by $\Delta_\theta$.

In actual numerical calculations, we specify energy eigenstates corresponding to $V_\theta$ and $i \delphi$
by using quantum numbers associated to each eigenstate.
When the size of the system $L$ is a multiple of 4,
the ground state of the Hamiltonian is in the sector of
vanishing total magnetization $S_z^\mr{tot} = 0$ and momentum $k=0$ whereas
the primary state $e^{i\theta}$  is the ground state of the sector of
$S_z^\mr{tot} = 1 $ and $k=\pi$. Similarly, $ i\delphi $ state is the ground state
of the sector of $S_z^\mr{tot} =0$ and $k=2\pi/L$.

\subsubsection{Relative entropies: $ S( \rho_{V_\theta} \| \rho_\mr{GS})$,
$ S( \rho_{ i\delphi } \| \rho_\mr{GS})$,  $S( \rho_{i \delphi} \| \rho_{V_\theta})$.  } 
\begin{figure*}
 \includegraphics[width = 8.5cm]{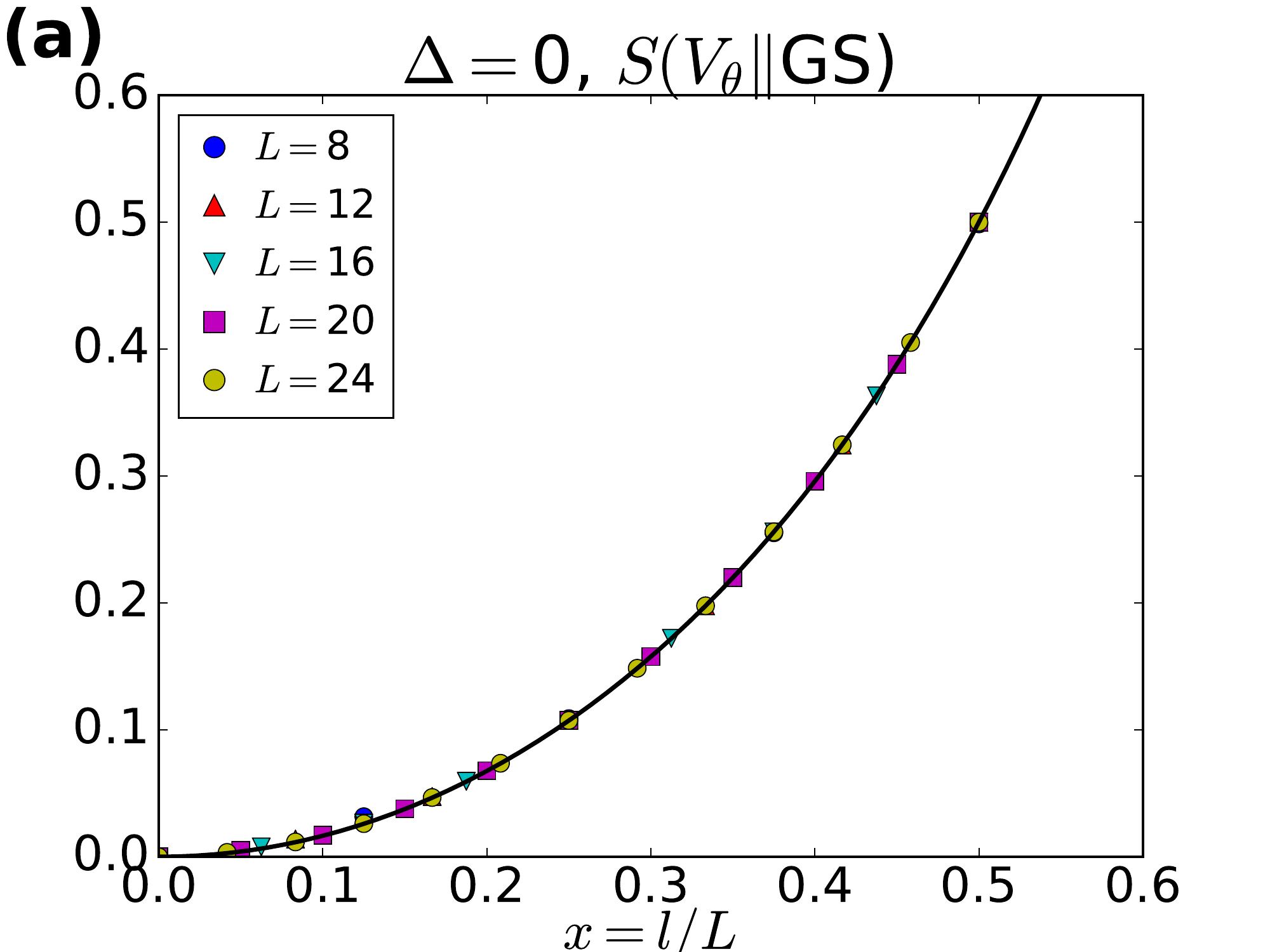}
 \includegraphics[width = 8.5cm]{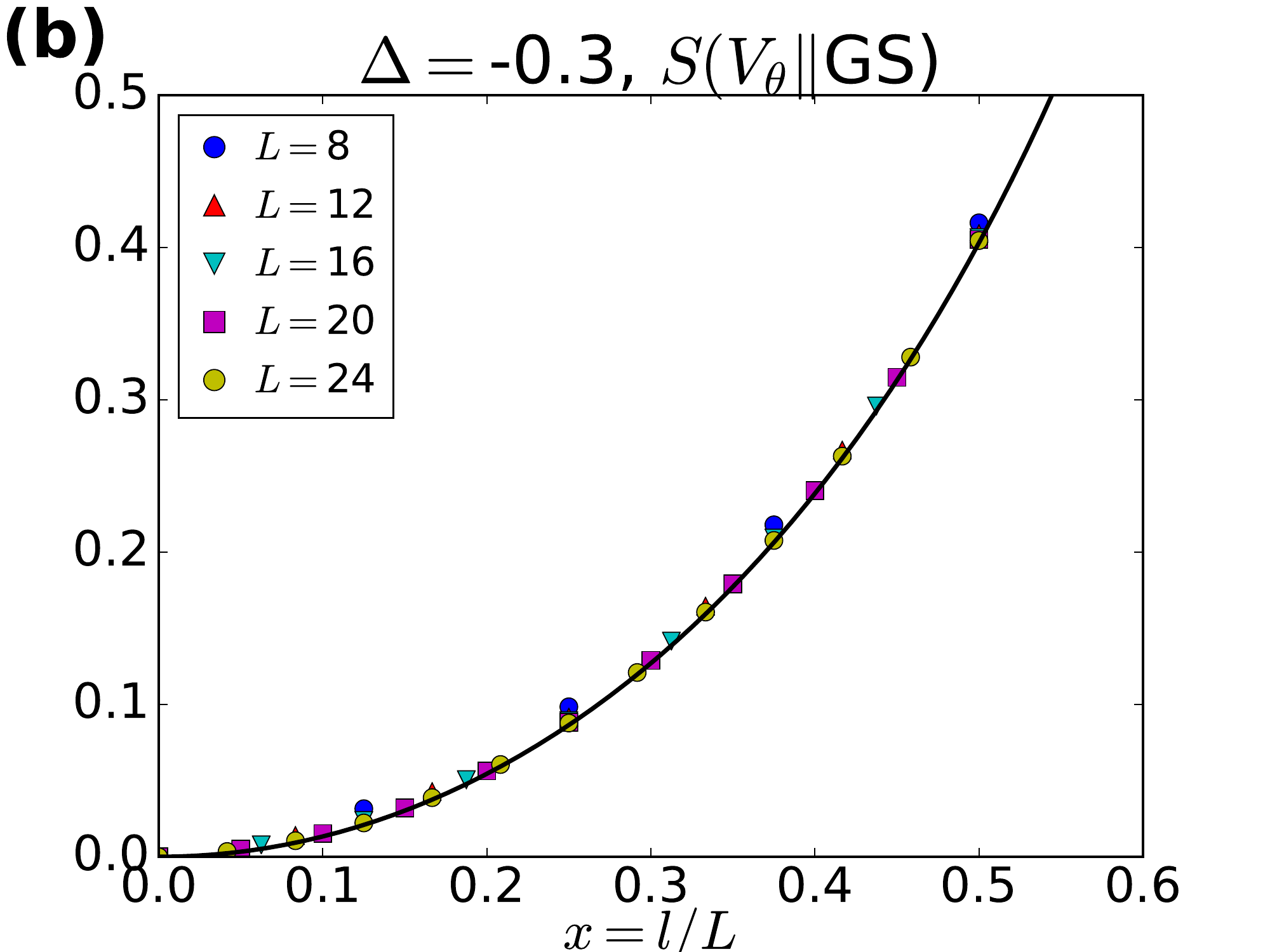}
 \includegraphics[width = 8.5cm]{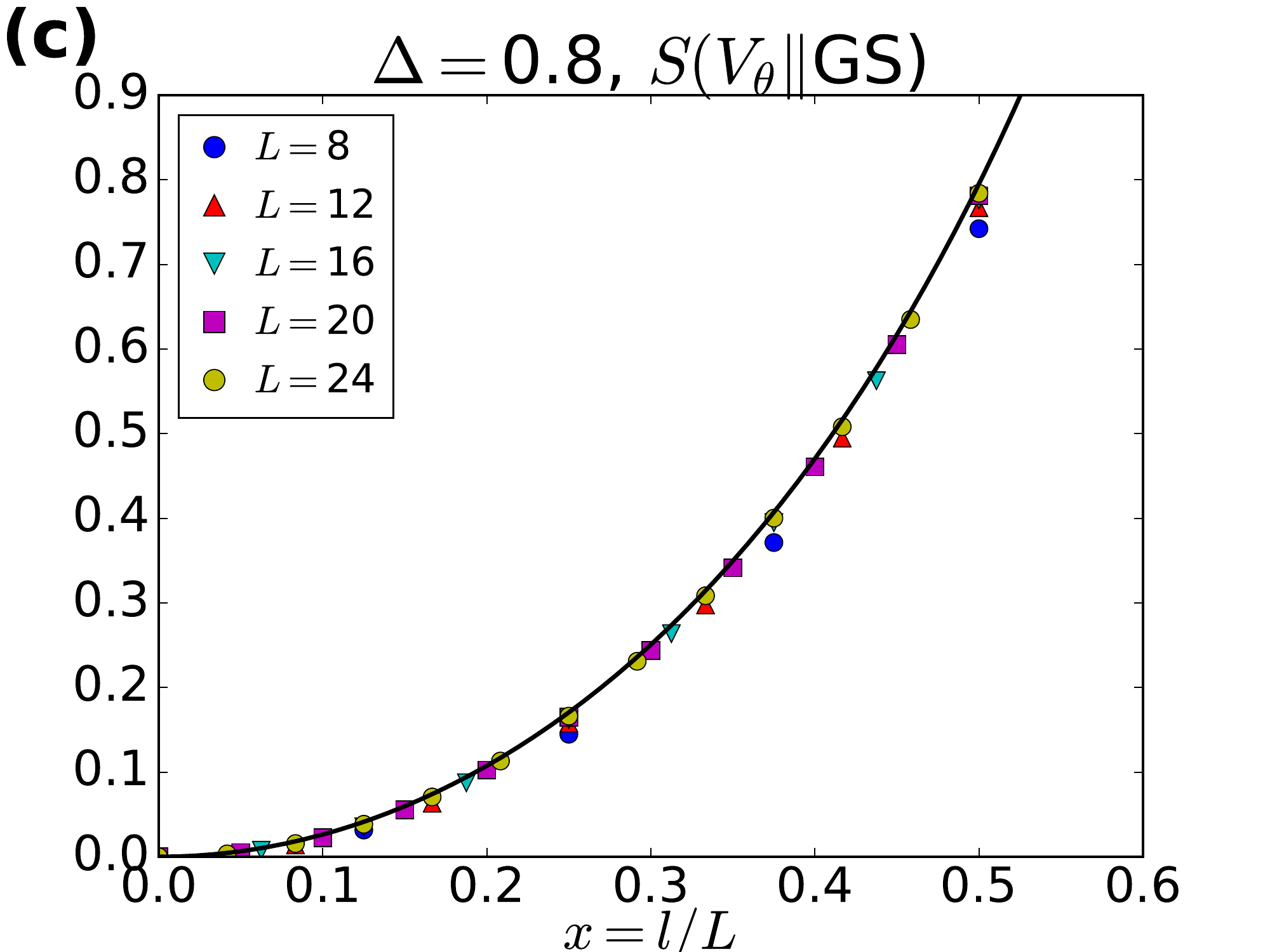}
 \includegraphics[width = 8.5cm]{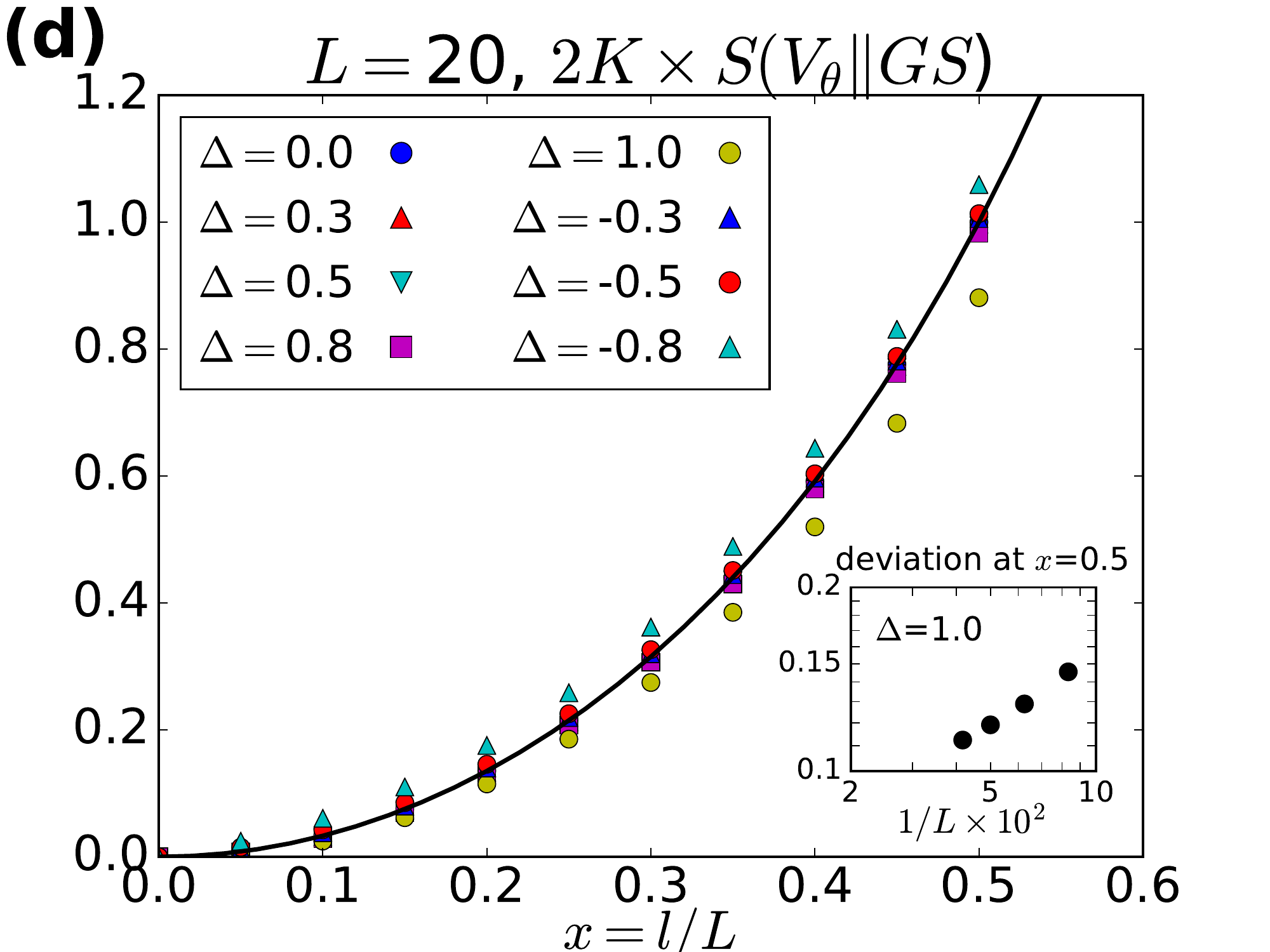}
 \caption{(a)(b)(c) Numerical results of the relative entropy between $V_\theta$ and the ground state of
 the XXZ chain for several values of $\Delta$ and $L$.
 Lines are the CFT prediction~\eref{formula_general_V_gs} with $\Delta_V = 1/(4K)$.
 (d) Plot of $2K \times S( \rho_{V_\theta} \| \rho_\mr{GS}) $, which is expected to
 be independent of $\Delta$ from the CFT prediction, $1-\pi x \cot(\pi x)$ (line).
 The inset shows the deviations of the numerical
  data of $\Delta = 1.0$ from the CFT prediction at $x=0.5$.}
 \label{Fig_V1_GS}
\end{figure*}
\begin{figure*}
 \includegraphics[width = 8.5cm]{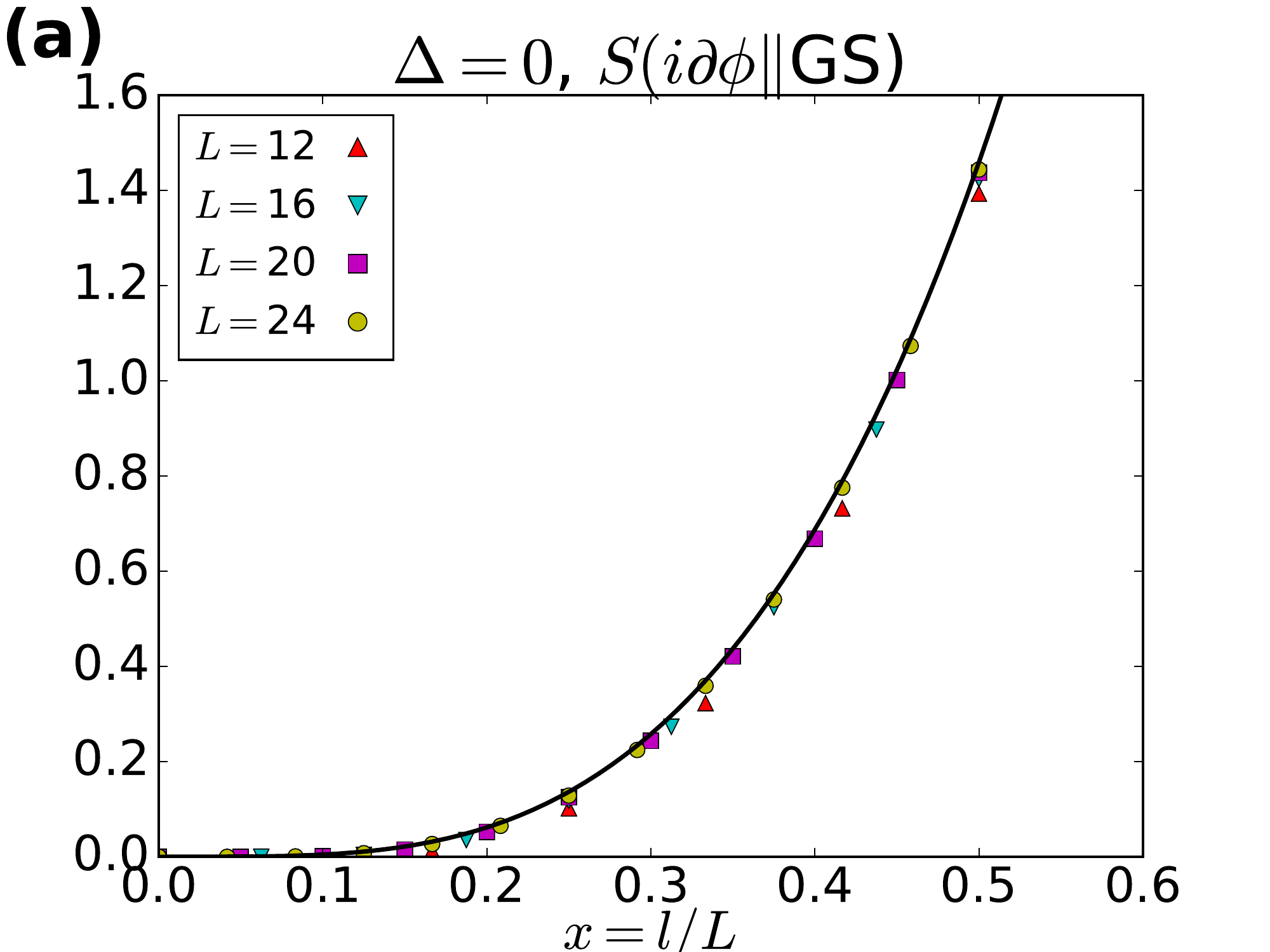}
 \includegraphics[width = 8.5cm]{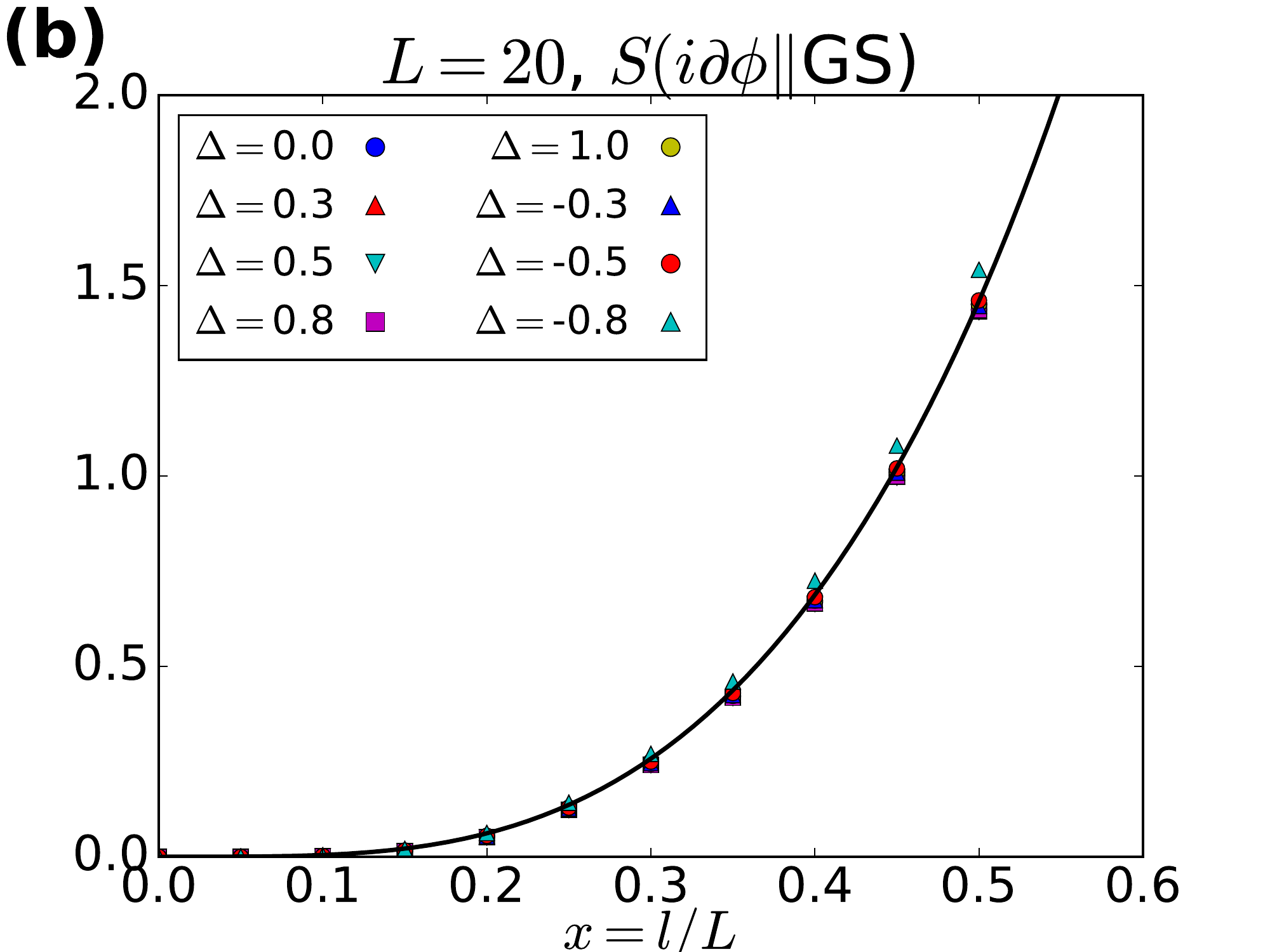}
 \caption{Numerical results of the relative entropy between $i \delphi $ and
 the ground state of the XXZ chain for (a) several $L$ with $\Delta=0$ and
 (b) several $\Delta$ with $L=20$. Lines are the CFT prediction~\eref{formula_delphi_gs}. }
 \label{Fig_delphi_GS}
\end{figure*}
\begin{figure}
 \includegraphics[width = 8.5cm]{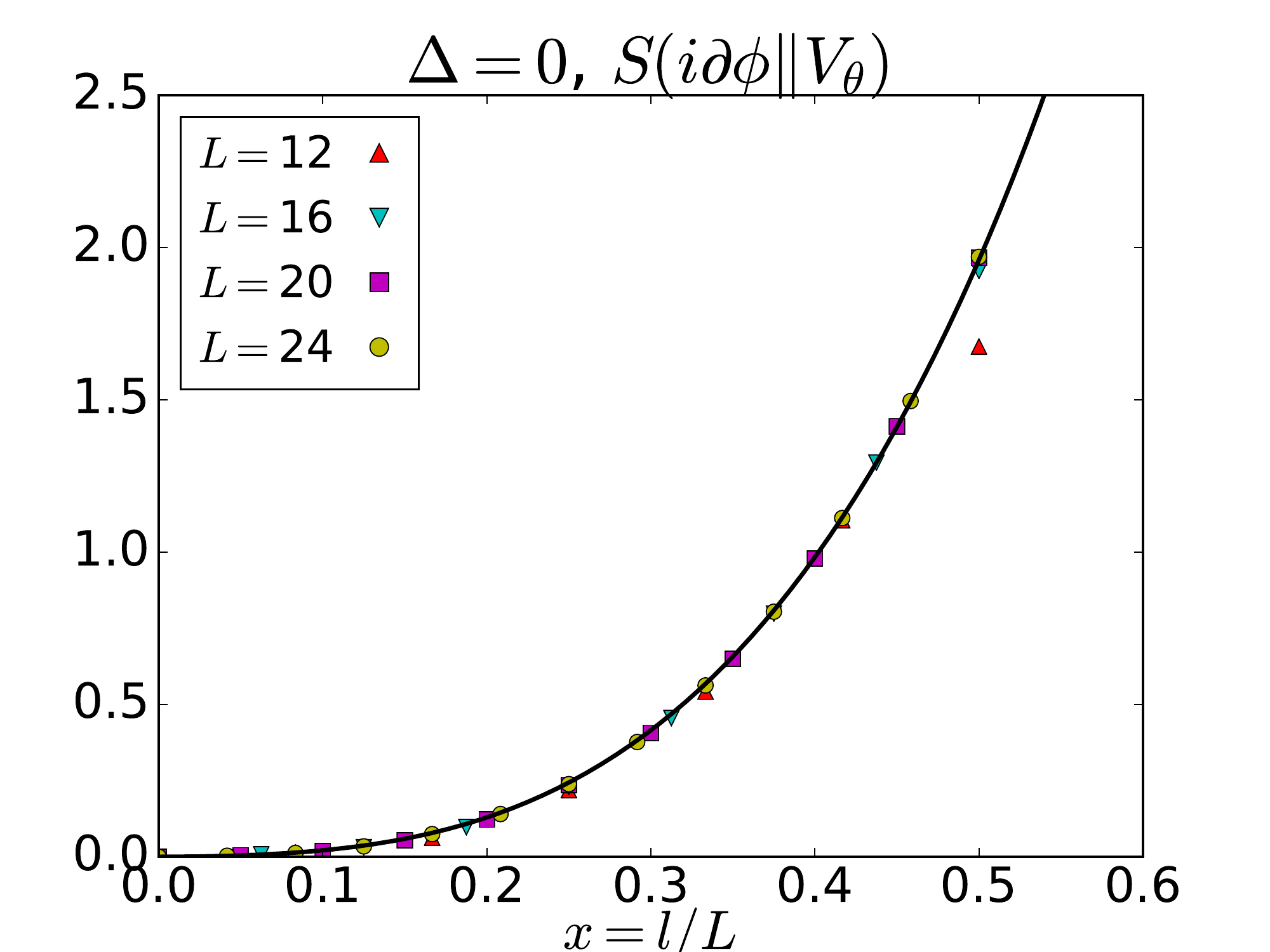}
 \caption{Numerical results of the relative entropy between the $i \delphi $ and $V_\theta$ of the XXZ chain
 at $\Delta = 0$. Line is the CFT prediction~\eref{relation_delphi_vertex}. }
 \label{Fig_delphi_V1}
\end{figure}

In Fig.~\ref{Fig_V1_GS}, we show the numerical results of $ S( \rho_{V_\theta} \| \rho_\mr{GS})$
for several values of $\Delta$ and $L$.
The numerical data perfectly agree with
the CFT predictions~(\eref{formula_general_V_gs} with $\Delta_V = 1/(4K)$ and
note that $K$ depends on $\Delta$).
We observe that the finite size effect (deviation from the CFT predictions) is stronger for larger $|\Delta|$
(see Fig.~\ref{Fig_V1_GS}(d)).
We think this is a consequence of the irrelevant term coming from $\Delta \sum_i S_i^z S_{i+1}^z$
in the derivation of the continuum field theory~\eref{freeboson_action}
from the lattice model~\eref{XXZ_model}~\cite{GiamarchiBook}.
\footnote{We note that the relatively large deviations from the CFT predictions in the case of $\Delta=1$
are possibly due to the logarithmic corrections such as $\sim 1/(\ln L)$ coming from
the marginally irrelevant operator~\cite{Cardy_log_correction}.}

Next, in Fig.~\ref{Fig_delphi_GS}, we present the numerical results of $S( \rho_{i \delphi} \| \rho_\mr{GS})$.
Agreement between the numerical data and the CFT prediction is again quite well.

Finally, we show the numerical data of $ S( \rho_{i \delphi} \| \rho_{V_\theta} )$ in Fig.~\ref{Fig_delphi_V1}.
We confirm the CFT prediction~\eref{relation_delphi_vertex},
$ S( \rho_{i \delphi} \| \rho_{V_\theta} ) = S( \rho_{i \delphi} \| \rho_\mr{GS}) + S( \rho_{V_\theta} \| \rho_\mr{GS})$. \footnote{ Note that $S( \rho_{V_\theta} \| \rho_\mr{GS}) =S( \rho_\mr{GS}\|  \rho_{V_\theta})$ holds
in the case of $c=1$ CFT~\cite{Ruggiero:2016khg}.}

\subsubsection{Sandwiched R\'{e}nyi relative entropy.}
\begin{figure*}
 \includegraphics[width = 8.5cm]{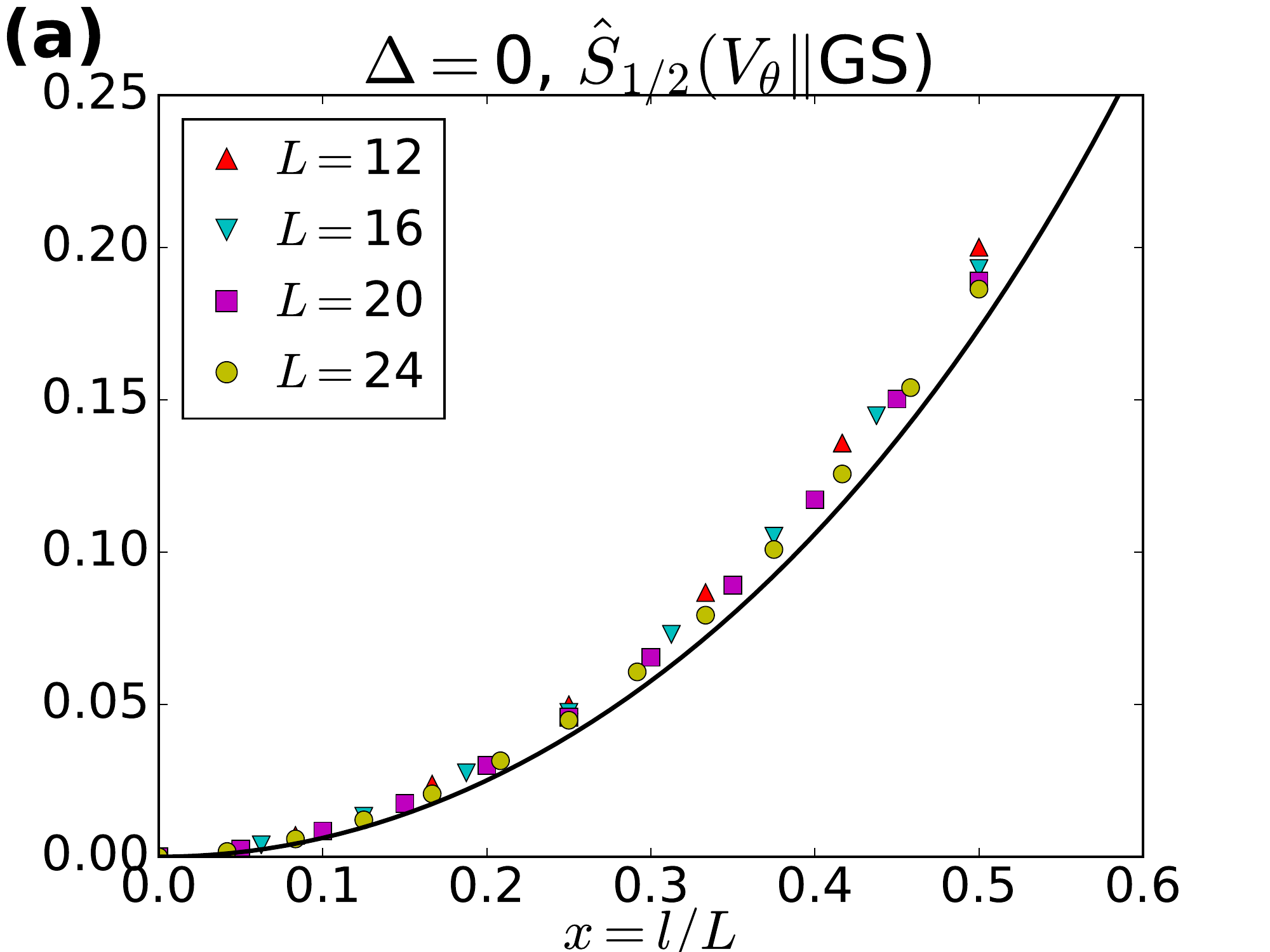}
 \includegraphics[width = 8.5cm]{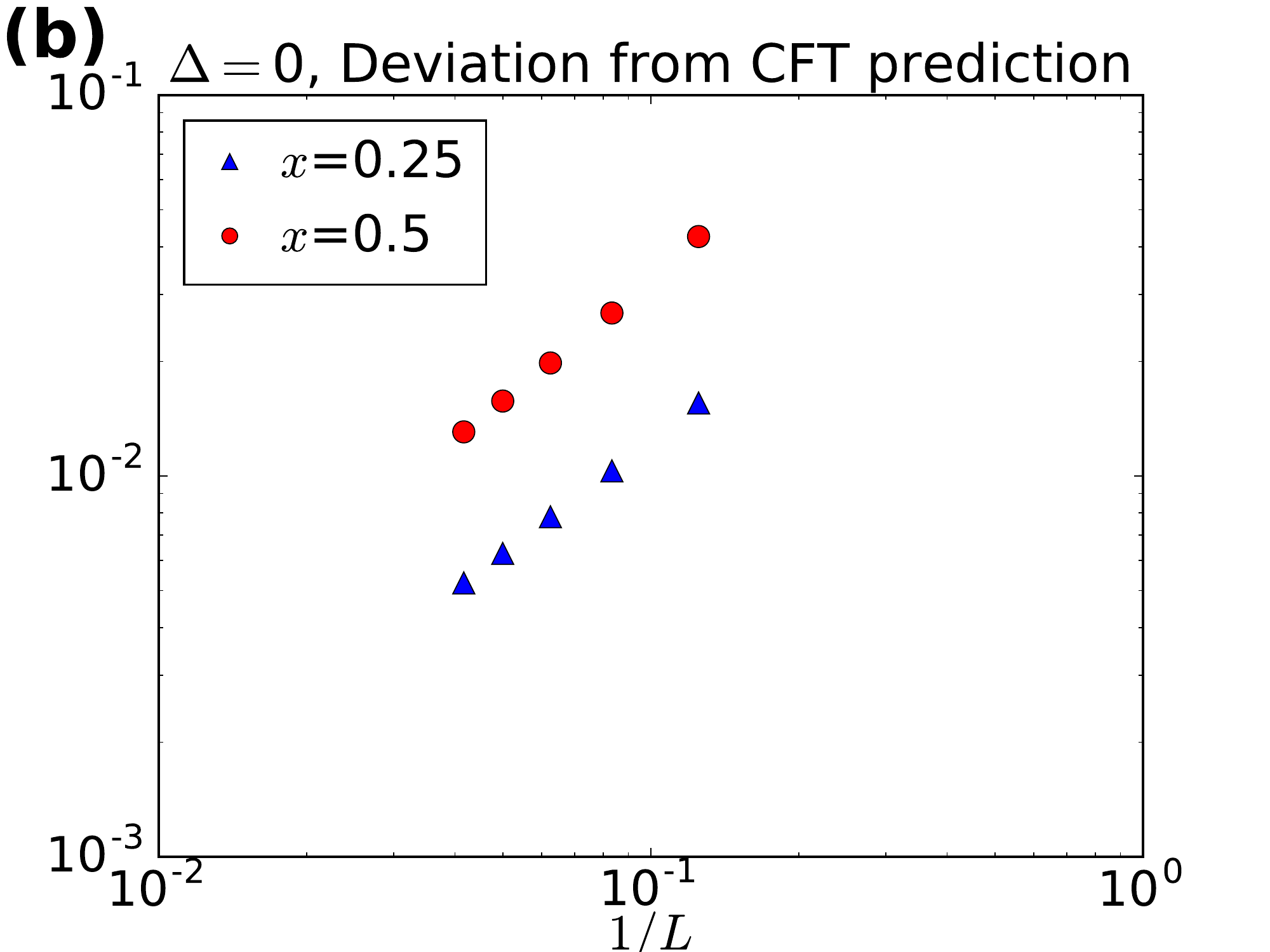}
 \caption{(a) Numerical results of the sandwiched R\'enyi relative entropy of the order of $\alpha=1/2$
 between $V_\theta$ and the ground state of the XXZ chain ($\hat{S}_{1/2}(\rho_{V_\theta} \| \rho_\mr{GS})$).
 Line is the CFT prediction~\eref{formula_fidelity}.
 (b) The deviation of the numerical data from the CFT prediction at $x=0.25$ and $x=0.5$.}
 \label{Fig_Fidelity_V1_GS}
\end{figure*}

As mentioned in the introduction, the meaning of the R\'enyi relative entropy~\eref{Def_RenyiRelEE},
if any, has not been revealed yet from the viewpoint of quantum information theory. 
Here we numerically calculate the sandwiched R\'enyi relative entropy~\eref{eq:sand}
which was shown to have desirable properties in quantum information theory~\cite{2013JMP....54l2201F}  
by taking advantage of exact diagonalization.

In~\cite{Lashkari:2014yva} Lashkari derived a formula of 
the sandwiched R\'enyi relative entropy of the order of $\alpha=1/2$
between the vertex operator and the ground state in $c=1$ CFT and the result leads to
\be
 \hat{S}_{1/2}(\rho_{V_\theta} \| \rho_\mr{GS}) = - 2 \Delta_\theta \log \lp \cos(\pi x/2) \rp. \label{formula_fidelity}
\ee
We numerically calculate $\hat{S}_{1/2}(\rho_{V_\theta} \| \rho_\mr{GS})$ in the XXZ chain 
and compare it with the prediction by CFT (Fig.~\ref{Fig_Fidelity_V1_GS}).
We observe that the numerical data match the CFT prediction
although the finite size effect is larger than that of the relative entropies calculated in the above.

\subsection{Critical transverse field Ising chain: the $c=1/2$ Ising CFT}
\begin{figure*}
 \includegraphics[width = 8.5cm]{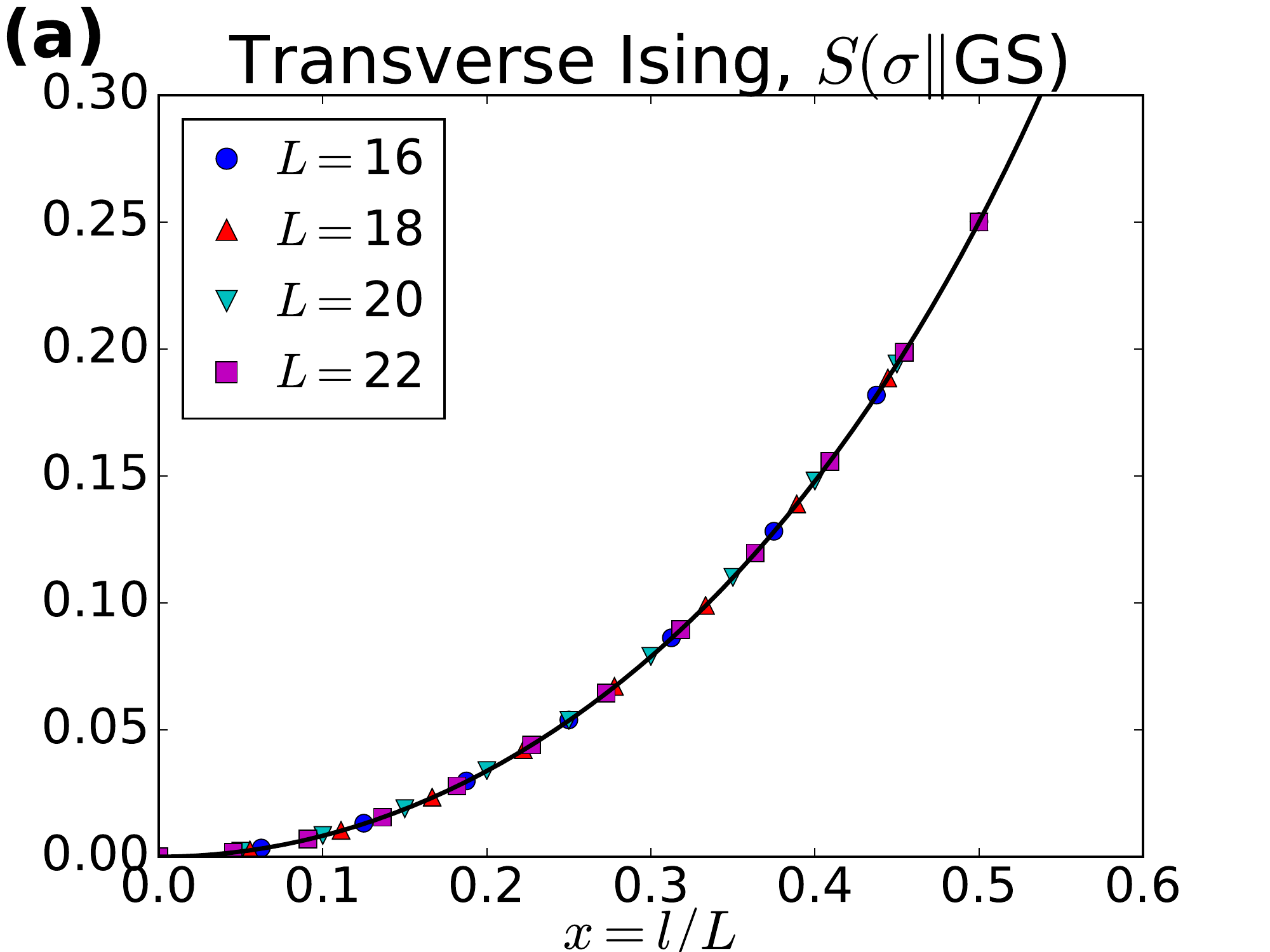}
 \includegraphics[width = 8.5cm]{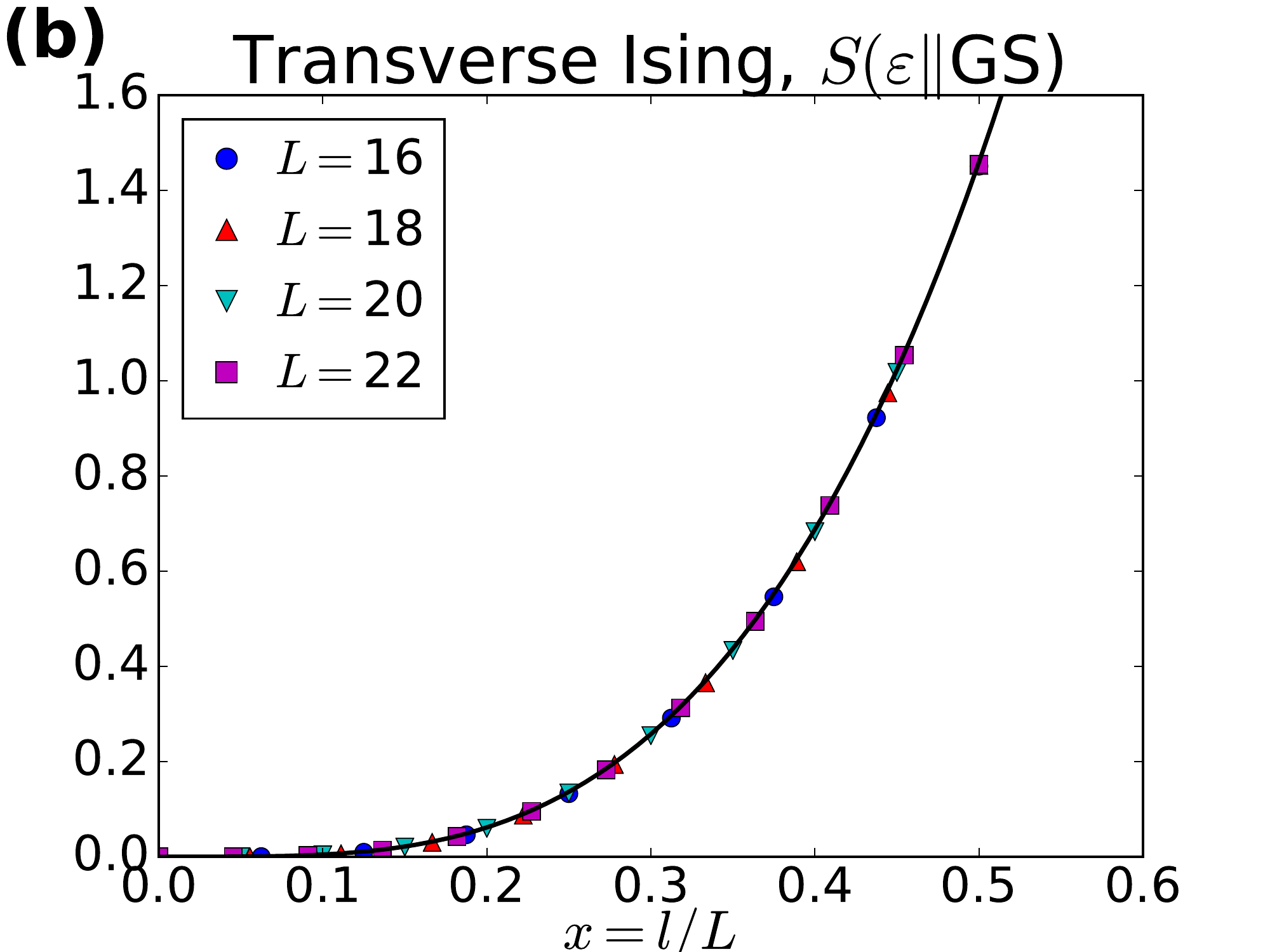}
 \includegraphics[width = 8.5cm]{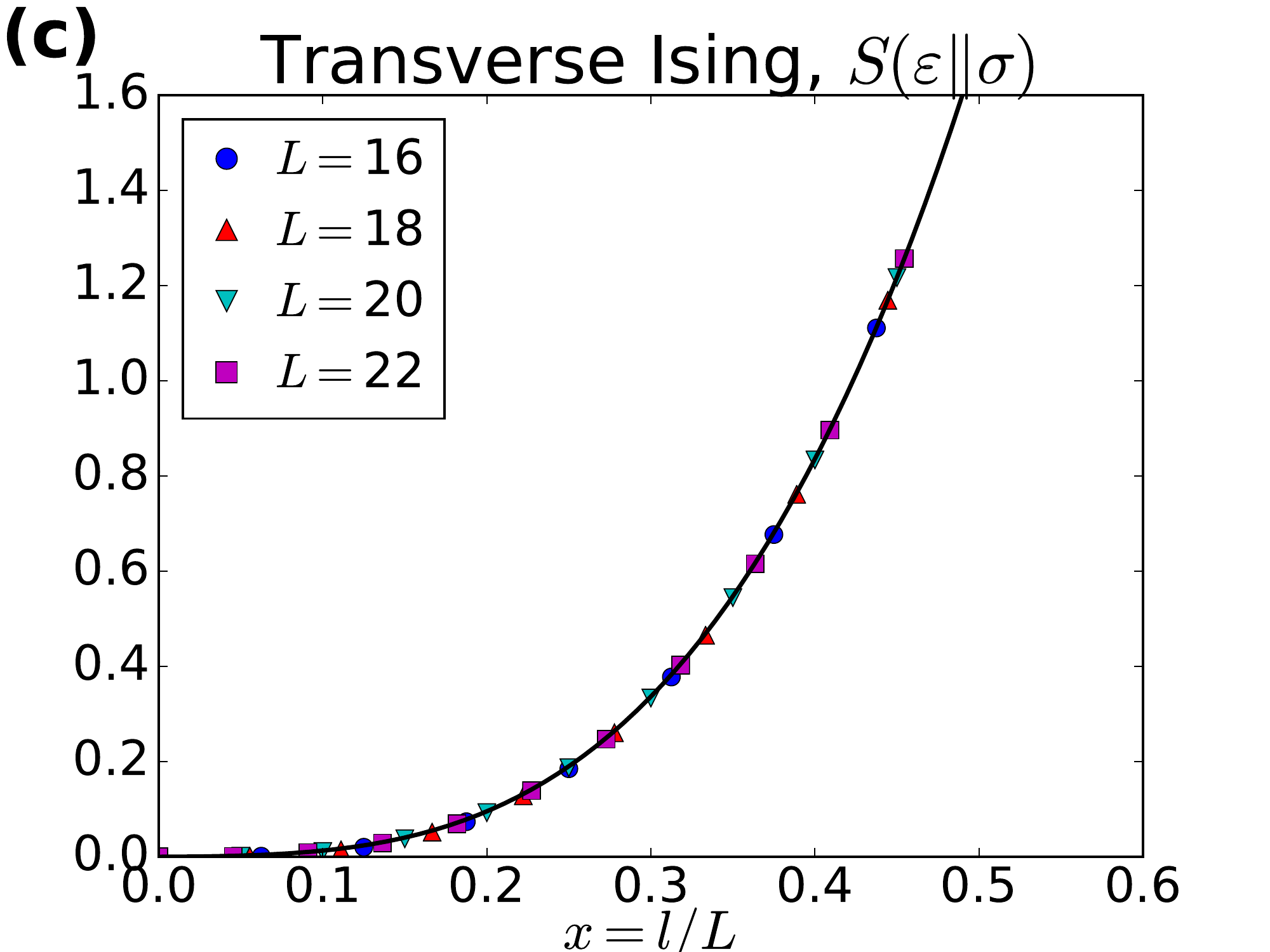}
 \caption{Numerical results of the relative entropy
 among the primary states $\sigma, \varepsilon$, and the ground state
 of the critical transverse field Ising chain: (a) $S(\rho_{\sigma} \| \rho_\mr{GS})$,
 (b) $S(\rho_{\varepsilon} \| \rho_\mr{GS})$, and (c) $S(\rho_{\varepsilon} \| \rho_{\sigma})$. 
 Lines are the CFT predictions~\eref{TI_sigma_gs}, \eref{TI_epsilon_GS}, \eref{TI_epsilon_sigma}
 (note that $S(\rho_{\sigma} \| \rho_\mr{GS}) =  S(\rho_\mr{GS} \| \rho_{\sigma})$ holds again
 as in the case of $c=1$ CFT~\cite{Ruggiero:2016khg}).  }
 \label{Fig_TI}
\end{figure*}

The next example where we compare numerical results of a lattice model with CFT predictions
is the critical transverse field Ising chain
under periodic boundary condition,
\begin{equation}
 H = \sum_{i=1}^L \left( S^x_i S^x_{i+1} + 0.5 S^z_i  \right).  \label{TI_model}
\end{equation}
The low-energy physics of this model is described by the $c=1/2$ Ising CFT~\eref{c=1/2: action}.
We perform exact diagonalization of this model and calculate the relative entropies
between the eigenstates corresponding the primary states $\sigma, \varepsilon$
in the same way as the XXZ chain.
Specifically, when $L$ is even, the ground state is in the sector of the momentum $k=0$ 
while the primary state $\sigma$ ($\varepsilon$) is obtained as
the ground state (the first excited state) of the sector of the momentum $k=\pi \: (0)$.

In Fig.~\ref{Fig_TI}, the numerical results for
$S(\rho_{\sigma} \| \rho_\mr{GS}), S(\rho_{\varepsilon} \| \rho_\mr{GS}) $,
and  $S(\rho_{\varepsilon} \| \rho_{\sigma}) $ are presented.
The agreement between the CFT predictions and the numerical data is remarkably well.

\section{Discussions and outlook}
\label{section:conc}
In this work we investigated the relative entropies between several primary states in the critical spin chains
and compared them with the predictions by corresponding CFTs.
After reviewing the results of~\cite{Ruggiero:2016khg} for $c=1$ free boson CFT
we analytically derived the formulae of the relative entropy in the $c=1/2$ Ising CFT.
Then we numerically calculated the relative entropy between several eigenstates
in two critical spin chains, the $S=1/2$ XXZ chain and the critical transverse field Ising chain,
by exact diagonalization. 
We saw perfect agreements of the numerical results with the CFT predictions in both models.
Our numerical results extend those of~\cite{Ruggiero:2016khg}
to the relative entropy itself rather than the R\'enyi counterpart of it and to interacting lattice systems. 
Our results establish further confirmation on the correspondence of the low-energy physics and 
the structure of entanglement between critical lattice systems and appropriate CFTs. 

As a future direction, it would be interesting to extend our results
to models corresponding to more complicated (1+1)-dimensional CFTs such as the minimal CFTs
of $c=7/10, 4/5, \cdots$ or level-$k$ Wess-Zumino-Witten model.
Another direction of interest is to investigate the relative entropy in higher dimensions
where little is known in the literature~\cite{Sarosi:2016atx}.

\section*{Acknowledgements}
Y.O.N. acknowledges T. Giamarchi, K. Okamoto, K. Damle, and M. Oshikawa for valuable discussions.
Y.O.N. was supported by Advanced Leading Graduate Course for Photon Science (ALPS)
of the Japan Society for the Promotion of Science (JSPS) and by JSPS KAKENHI Grants No. JP16J01135.
Y.O.N. thanks the hospitality of Kavli Institute for Theoretical Physics, where this work was initiated. 
The work of T.U. was supported in part by the National Science Foundation under Grant
No. NSF PHY-1125915.

\section*{References}
\bibliographystyle{utphys}
\bibliography{relative}

\end{document}